\documentclass{article}
\usepackage{ijcai26}

\usepackage{times}
\usepackage{soul}
\usepackage{url}
\usepackage[hidelinks]{hyperref}
\usepackage[utf8]{inputenc}
\usepackage[small]{caption}
\usepackage{graphicx}
\usepackage{amsmath}
\usepackage{amsthm}
\usepackage{booktabs}
\usepackage{algorithm}
\usepackage{algorithmic}
\usepackage[switch]{lineno}
\usepackage[table]{xcolor}      
\usepackage{multirow}      
\usepackage{pifont}

\title{PicoSpec: A Pipelined Collaborative Speculative Decoding Framework for Efficient Edge-Cloud LLM Inference}

\author{
Yida Zhang$^{1}$\thanks{Equal contribution.}
\and
Zhiyong Gao$^{1}$\footnotemark[1]
\and
Shuaibing Yue$^1$
\and
Jie Li$^1$
\and
Rui Wang$^{1}$\thanks{Corresponding author.}
\\
\affiliations
$^1$University of Science and Technology Beijing, School of Computer and Communication Engineering\\
\emails
ustbzyd@163.com,
gzy@xs.ustb.edu.cn,
m202421086@xs.ustb.edu.cn,
m202521017@xs.ustb.edu.cn,
wangrui@ustb.edu.cn
}

\begin{document}

\maketitle

\begin{abstract}
Recent advancements and widespread adoption of Large Language Models (LLMs) in both industry and academia have catalyzed significant demand for LLM serving. However, traditional cloud services incur high costs, while on-device inference alone faces challenges due to limited resources. Edge-cloud collaboration emerges as a key research direction to combine the strengths of both paradigms, yet efficiently utilizing limited network bandwidth while fully leveraging and balancing the computational capabilities of edge devices and the cloud remains an open problem. To address these challenges, we propose \underline{\textbf{Pi}}pelined \underline{\textbf{Co}}llaborative \underline{\textbf{Spec}}ulative Decoding Framework (PicoSpec), a novel, general-purpose, and training-free speculative decoding framework for LLM edge-cloud collaborative inference. We design an asynchronous pipeline that resolves the mutual waiting problem inherent in vanilla speculative decoding within edge collaboration scenarios, which concurrently executes a Small Language Model (SLM) on the edge device and a LLM in the cloud. Meanwhile, to mitigate the significant communication latency caused by transmitting vocabulary distributions, we introduce separate rejection sampling with sparse compression, which completes the rejection sampling with only a one-time cost of transmitting the compressed vocabulary. Experimental results demonstrate that our solution outperforms baseline and existing methods, achieving up to 2.9× speedup.
\end{abstract}

\section{Introduction}
Large Language Models (LLMs) have shown incredible potential in tasks like natural language processing, coding, and reasoning. However, their massive memory and compute requirements make it difficult to deploy on resource-constrained edge devices. Edge-cloud collaborative inference offers a way out by offloading heavy work to the cloud while keeping easy task local\cite{10.1145/3719664}.

\begin{figure}[t]
    \centering
    \includegraphics[width=0.44\textwidth]{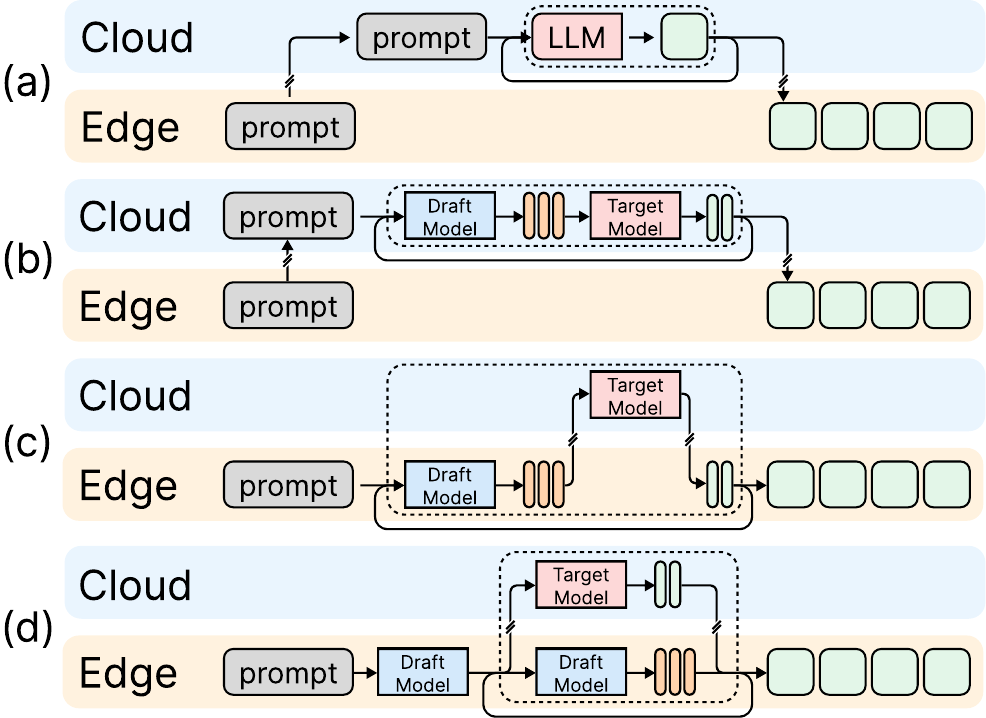}
    \caption{Comparison between (a) Cloud autoregressive decoding, (b) Cloud speculative decoding, (c) Vanilla collaborative speculative decoding, and (d) PicoSpec.}
    \label{fig:comp}
\end{figure}

Speculative decoding has proven to be an effective technique for improving inference speed\cite{chen2023accelerating}. Its core idea is to use a small "Draft" model to quickly generate draft tokens, then verified by a larger "Target" model (Figure \ref{fig:comp}b). Originally, this approach was designed to speed up the generation process on high-performance cloud servers, where data transfer is almost instantaneous. Someone tried to apply it to edge-cloud collaborative setups (Figure \ref{fig:comp}c). However, in practice, this transition is far from simple. When moved from a high-speed local environment to a distributed network, researchers have found that the original design faces massive performance drops and practical hurdles that make it difficult to use in the real world\cite{wang2024comprehensive}.

Despite the promise of edge-cloud collaborative speculative decoding, existing solutions still face severe "communication-computation mismatch" and flexibility issues in the real-world. Existing solutions use a serial "stop-and-wait" mechanism where the edge must pause for cloud verification before generating the next batch \cite{yu2025dsd,ning2025dssd}. In a Wide Area Network (WAN), the Round-Trip Time (RTT) is often high enough to cancel out any speedups, leaving the pipeline full of "bubbles". Other approaches suggest using early-exit points in the cloud to reduce verification latency\cite{venkatesha2025fast}. However, these invasive methods require model retraining or fine-tuning, which undermines generality and increases operational costs\cite{wang2024comprehensive}.

To solve these problems, we propose PicoSpec (as in Figure \ref{fig:comp}d), a plug-and-play framework designed for efficient distributed inference. PicoSpec is a system-level solution that works with standard models without any retraining. Its core consists of an asynchronous pipeline that decouples the edge-side drafting and cloud-side verification steps, as well as a separate rejection sampling algorithm optimized for communication overhead. This allows the edge device to keep speculating future tokens without waiting for the cloud's feedback on previous ones, effectively hiding the network latency. Our main contributions are as follows:

\begin{itemize}
\item PicoSpec is the first research to propose a training-free asynchronous framework for distributed speculative inference. We have comprehensively designed a generative inference framework for collaborative speculative decoding of LLMs, enabling faster inference without any modification or retraining of existing models.
\item PicoSpec introduces an asynchronous distributed speculative decoding pipeline. We incorporate Parallel Drafting, Fast Verification, and Overlapped Communication into the pipeline, which eliminates data dependencies between the edge device and the cloud, enabling genuine parallel decoding between edge and cloud.
\item PicoSpec designs Separate Rejection Sampling algorithm to minimize data transmission overhead. Since transmitting the full vocabulary distribution typically incurs substantial communication costs, we distribute the rejection sampling process between the edge and the cloud and apply sparse compression on distribute data, significantly reducing communication latency without altering sampling accuracy.
\end{itemize}

\section{Related Work}

\subsection{Speculative Decoding}

Speculative decoding accelerates LLM inference by substituting serial generation with a "draft-then-verify" paradigm. Chen et al.\cite{chen2023accelerating} demonstrated that a modified rejection sampling mechanism can guarantee the output remains mathematically identical to the target model while providing significant wall-clock speedups. To enhance draft quality and throughput, Medusa \cite{pmlr-v235-cai24b} employs multiple parallel decoding heads fine-tuned on a frozen backbone, while EAGLE \cite{10.5555/3692070.3693232} uses feature-layer extrapolation to generate high-quality drafts with minimal overhead. For recent parallel speculative decoding studies, PEARL\cite{liu2025pearl} tackles the mutual waiting problem between the draft model and target model by pre-verify and post-verify mechanisms, SwiftSpec \cite{zhang2025swiftspec} hides drafting costs via parallel tree generation, CoSine \cite{gao2025collaborative} scales this efficiency to multi-GPU environments by routing requests across specialized draft models.

While these methods perform well in data centers with high-speed interconnects (like NVLink). they struggle in high-latency Edge-Cloud environment. The main issues are limited computing power and the mismatch between communication costs and latency masking. There is still a lack of system-level design specifically for high-latency, low-bandwidth, and state-separated environments.

\subsection{Edge-Cloud Collaborative Inference}

Edge-Cloud collaborative inference via model spliting and task offloading to overcome the resource constraints of edge nodes. Frameworks like EdgeShard \cite{zhang2024edgeshard} and Jupiter \cite{ye2025jupiter} employ dynamic programming and intra-sequence pipeline parallelism to optimize model splitting across distributed nodes. However, these methods do not alter the auto-regressive nature of LLMs, leaving inference speeds heavily dependent on network RTT. Other research focuses on hybrid routing: MMSL \cite{ma2025multi} uses multi-stage scheduling, while Hybrid LLM\cite{ding2024hybrid} and Hybrid SLM and LLM \cite{hao2024hybrid} route requests or low-confidence tokens to the cloud based on task difficulty. Despite reducing local load, these schemes still suffer from stop-and-wait bottlenecks -- the edge must idle during remote processing.This idling ties inference speed to network RTT, canceling out offloading gains in high-latency environment. Consequently, system throughput is often capped by network physical limits rather than the actual hardware capacity.

\subsection{Edge-Cloud Collaborative Speculative Decoding}
To optimize edge performance, EdgeLLM \cite{xu2024edgellm} employs local parallel tree generation but imposes heavy memory and computing burdens on limited hardware. Recent collaborative efforts like Venkatesha et al. \cite{venkatesha2025fast} and HAT \cite{xie2025novel} introduce early exits or hidden-state exchanges, yet these often require invasive model modifications or suffer from high transmission overhead. While DSSD \cite{ning2025dssd}reduces uplink data through local re-sampling, it remains a serial "stop-and-wait" architecture strictly bound by network RTT. A more recent framework, DSD\cite{yu2025dsd} extends speculative decoding to distributed environments using an Adaptive Window Control (AWC) policy to tune speculation windows dynamically.Nevertheless, DSD primarily follows a serial stop-and-wait pattern Similarly, SLED \cite{li2025sled} focuses on multi-client throughput but fails to provide deep latency masking for individual users. These limitations underscore the need for an asynchronous, general-purpose system that decouples generation from verification while maintaining model generality and state consistency.

\section{Framework Design}

\subsection{Overall Framework} 
The PicoSpec framework aims to achieve efficient and fully parallelized collaborative speculative decoding in edge-cloud scenario, while simultaneously addressing the challenges posed by distributed communication.

\begin{figure}[htbp]
    \centering
    \includegraphics[width=\linewidth]{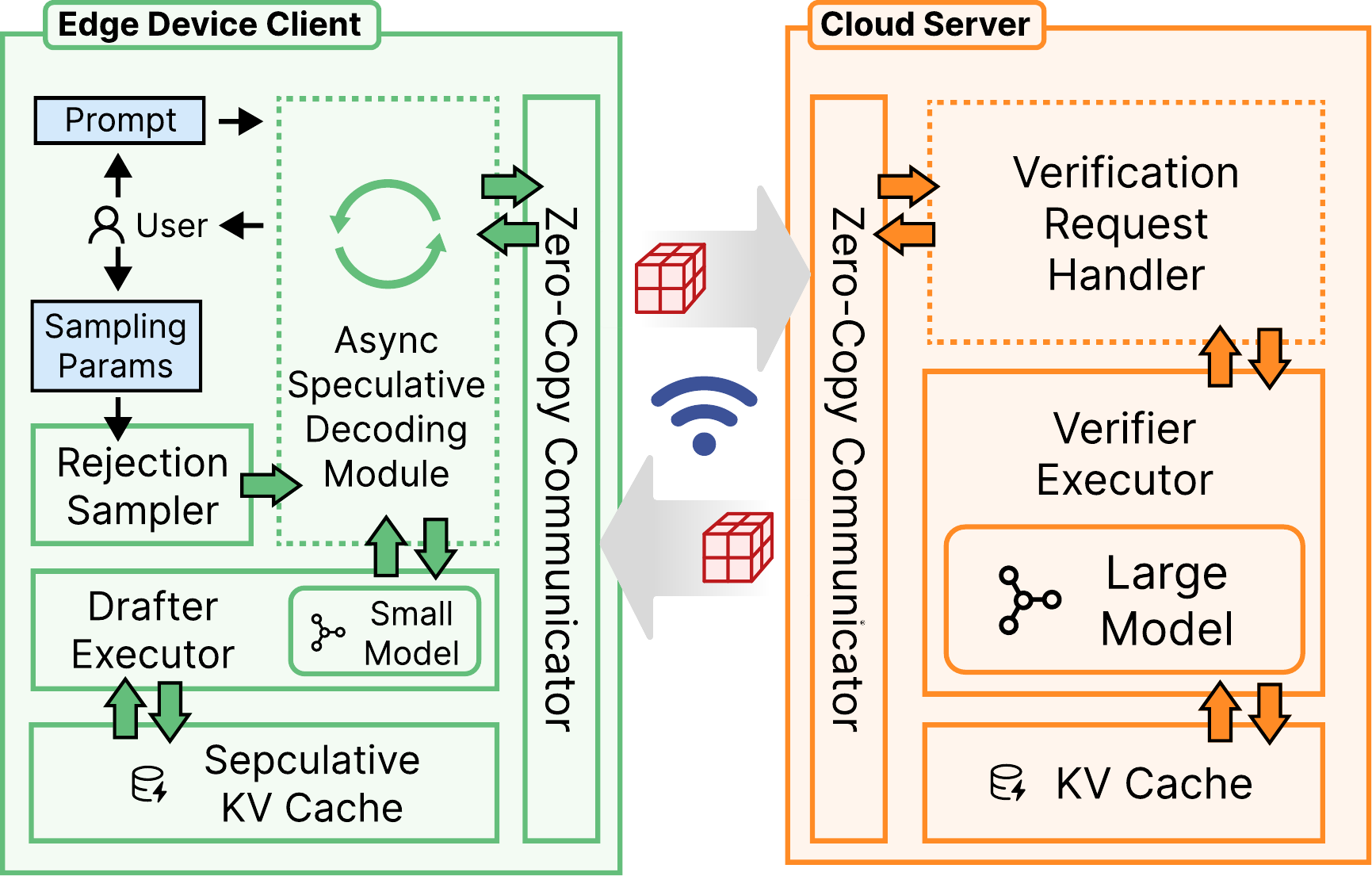}
    \caption{System Overview of PicoSpec Framework.}
    \label{fig:arch}
\end{figure}

The system overview of PicoSpec is illustrated in Figure \ref{fig:arch}.

Our framework comprises two main components: the edge-side and the cloud-side. On the edge side, it includes four modules: Parallel Drafter, Rejection Sampler, Speculative KV Cache, and Zero-Copy Communicator. On the cloud side, it consists of four modules: Verifier, Request Handler, KV Cache, and Zero-Copy Communicator.

After the user prepares the prompt and sampling parameters, both the edge and cloud simultaneously enter the prefill stage to initialize the request’s KV Cache. Subsequently, the user request proceeds to the collaborative speculative decoding stage. The edge-side Parallel Drafter module first executes drafting in parallel. Data is then transferred efficiently via the Zero-Copy Communicator module. The cloud-side Request Handler handles different types of verification requests. Finally, the Rejection Sampler module on the edge side completes the rejection sampling process to determine the final output. PicoSpec fully leverages KV cache to accelerate model execution. We also implemented an efficient rollback KV cache manager within the KV Cache modules on both the edge and cloud sides.

Decoupled design lays the foundation for parallelism, but there are still two main problems.  First, how to fully use parallelism, the edge needs to "speculate ahead" with draft generation while waiting for cloud verification, which requires precise scheduling to avoid pipeline bubbles and ensure execution continuity. To address this, we propose Parallel Drafting and Fast Verification in Section 3.2.

Second, the system faces a communication-overhead trap. In a WAN environment, even a small increase in message frequency or data size can quickly saturate the limited bandwidth and negate any gains from speculation. Efficiently pruning data and fusing requests without losing model accuracy is a delicate balancing act. We address these critical efficiency issues through the optimized sampling and communication techniques described in Section 3.3.

\begin{figure*}[htbp]
    \centering
    \includegraphics[width=0.92\textwidth]{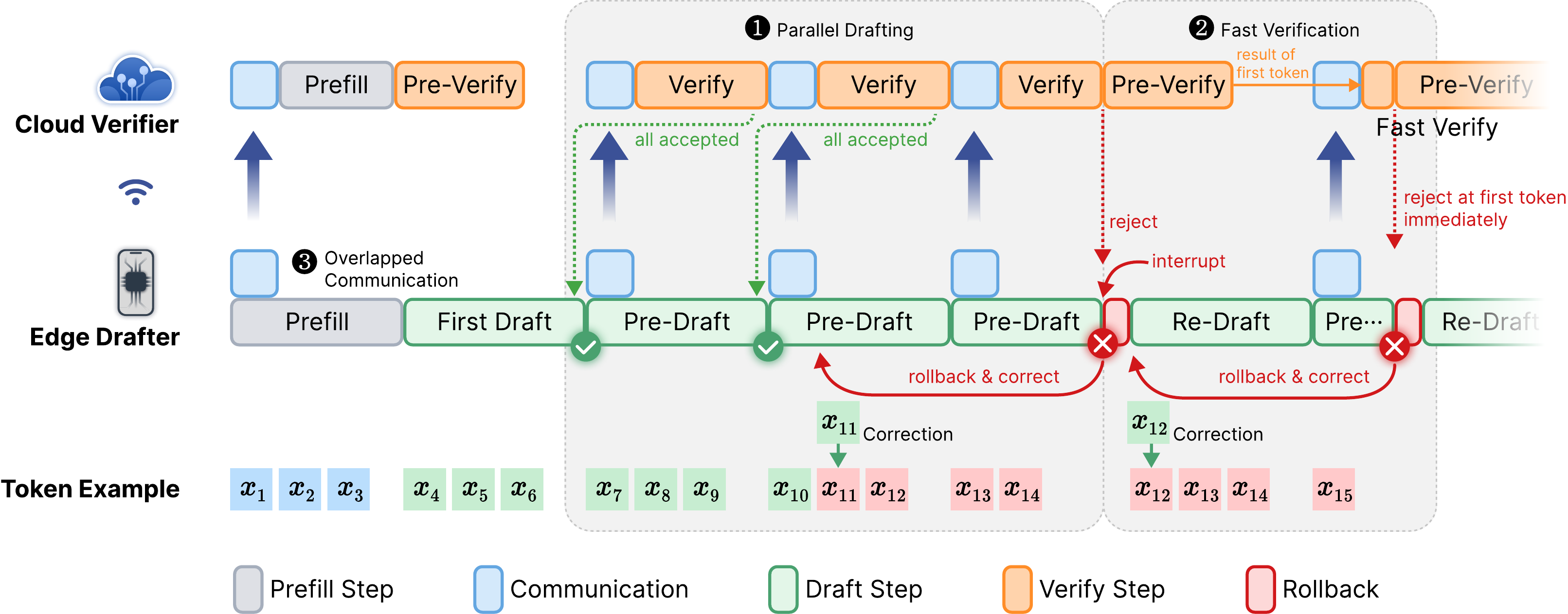}
    \caption{Asynchronous Pipeline of PicoSpec. \ding{182} \textbf{Parallel Drafting}: After the first draft, we perform multiple Pre-Draft steps consecutively. The i-th Pre-Draft on the edge is executed simultaneously with the (i-1)-th Verification in the cloud. \ding{183} \textbf{Fast Verification}: When a verification failure occurs, we immediately trigger a Pre-Verifiy step, and the subsequent Verify step quickly returns the verification result of the first token. \ding{184} \textbf{Overlapped Communication}: The sending/receiving of  data stream on the edge overlaps with Draft step.}
    \label{fig:pipeline}
\end{figure*}

The lifecycle of PicoSpec, shown as Figure \ref{fig:pipeline}, moves from a synchronous prefill to a continuous execution loop. After the initial prefill aligns the context between the edge and cloud, the cloud provides a seed token to trigger the speculation process. In the core Asynchronous Speculative Loop, the edge follows Parallel Drafting, drafting the next batch while the cloud verifies the previous one. To further cut down on waiting time, we use Fast Verification step that allows the cloud to start its work before the full draft even arrives. Meanwhile, we introduced Overlapped Communication, the data transmission time between edge-clouds is also successfully masked. If the cloud rejects a token, it immediately sends an interrupt signal to stop the edge's now-invalid drafting tasks. The system then performs a rollback and correction phase to restore state consistency before resuming the loop. 

\subsection{Edge-Cloud Collaborative Speculative Decoding}

PicoSpec decouple the local drafting logic from the remote verification process through asynchronous pipeline. In order to achieve better performance, achieving true edge collaborative speculative decoding. It is necessary to solve the data dependency problem at the edge and the cloud side.

In traditional synchronous stop-and-wait mode, the end-to-end latency for a single inference step, denoted as $L_{sync}$, is the linear sum of each phase's duration:

\begin{equation}
L_{\text{sync}} = T_{\text{draft}} + T_{\text{RTT}} + T_{\text{verify}}
\end{equation}

Here, $T_{draft}$ represents the total time for the edge to generate $\gamma$ draft tokens, $T_{RTT}$ is the network round-trip time, and $T_{verify}$ is the time required for parallel verification on the cloud. Due to strict data dependencies in this mode, the edge must keep idle during the $T_{RTT} + T_{verify}$ interval while waiting for the cloud to respond.

\begin{figure}[h]
    \centering
    \includegraphics[width=0.9\linewidth]{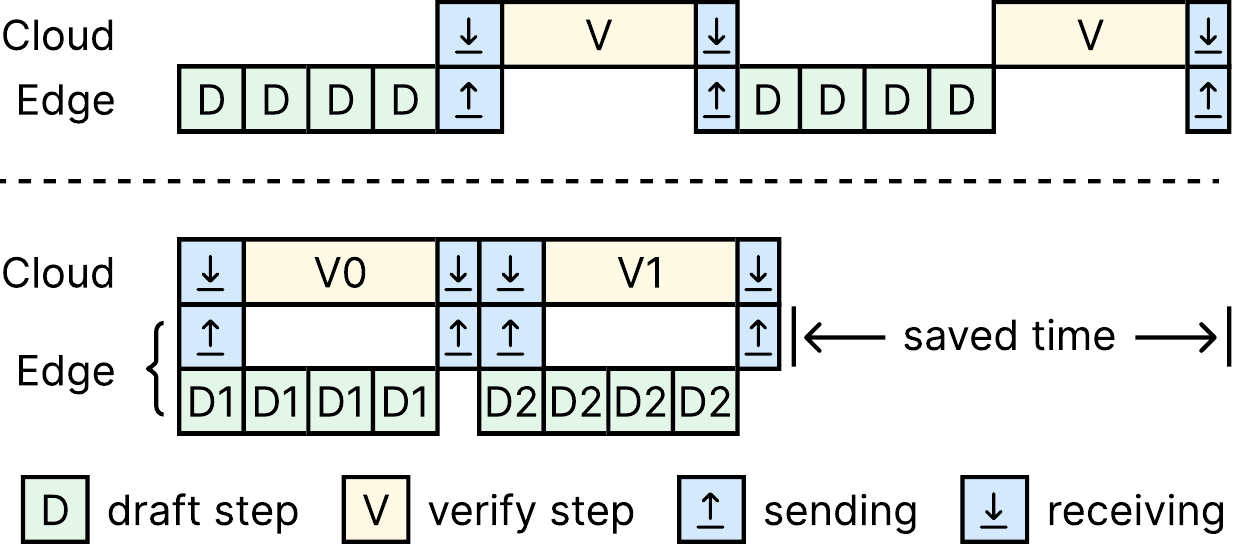}
    \caption{Parallel Drafting in asynchronous pipeline.}
    \label{fig:parallel_drafting} 
\end{figure}

As shown in Figure \ref{fig:parallel_drafting}, PicoSpec solves the data dependency problem of edge side by introducing Parallel Drafting. In the main generation loop, once the model execution thread completes drafting the current batch $S_i$ and submits a verification request, control is immediately returned to the system via a pre-draft. Instead of waiting for remote feedback, the main thread assumes $S_i$ is fully accepted. It uses the tail state of $S_i$ as the logical context to immediately start generating the next batch, $S_{i+1}$. Ideally, when $T_{draft} \ge T_{RTT} + T_{verify}$, the verification of step $i$ and the generation of step $i+1$ overlap completely on the timeline. Under these conditions, the amortized latency $L_{async}$ becomes:

\begin{equation}
    L_{async} = \max(T_{draft}, T_{RTT} + T_{verify})
\end{equation}

This design transforms the serial dependency chain of traditional cooperation speculative decoding into a parallel execution flow. As a result, system throughput is no longer bound by RTT and is instead limited only by the continuous productivity of the edge's draft generation.

\begin{figure}[h]
    \centering
    \includegraphics[width=0.9\linewidth]{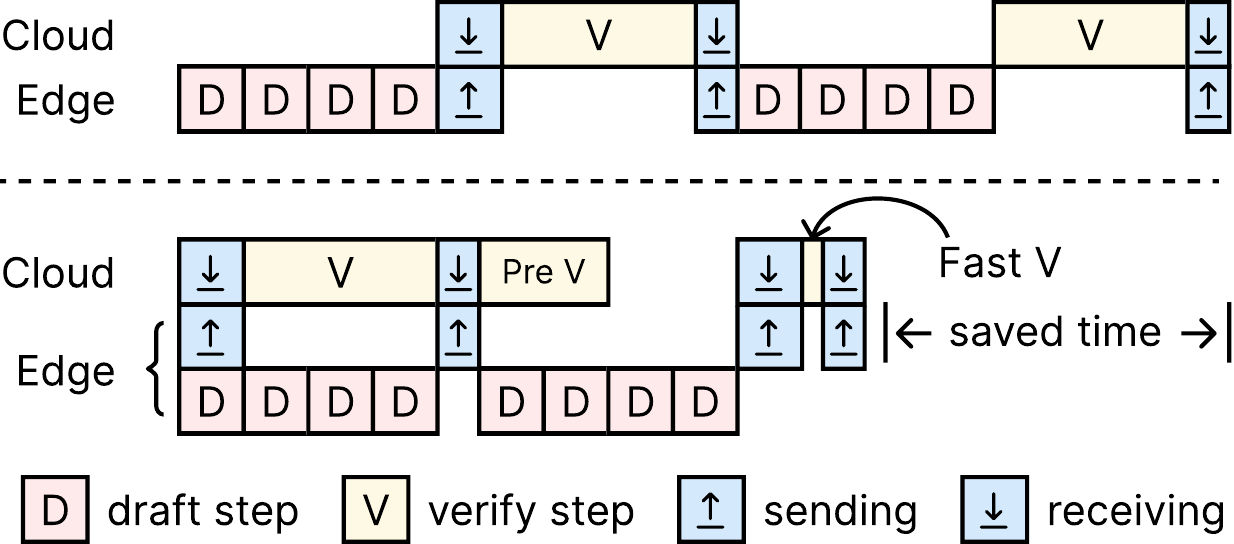}
    \caption{Fast Verification in asynchronous pipeline.}
    \label{fig:fast_verification} 
\end{figure}

To solve the problem of data dependence on the cloud side, PicoSpec propose Fast Verification mechanism. This mechanism aims for a secondary overlap of computation and communication at a sub-batch level. While standard asynchronous modes wait for the complete draft sequence $S_i$ to arrive at the cloud before starting verification, Fast Verification allows the PreVerify control signal as soon as partial tokens in $S_{i+1}$ are determined. Figure \ref{fig:fast_verification} illustrates the difference before and after the introduction of Fast Verification.

If we define the pre-verify lead time as $T_{pre}$, the effective start time of cloud verification, $T_{start}$, is moved forward. Consequently, the bubble time $T_{bubble}$ in the pipeline is further reduced to:

\begin{equation}
    T_{bubble} = \max(0, T_{RTT} + T_{verify} - T_{draft} - T_{pre})
\end{equation}

By using the powerful parallel forward-computation capabilities of the cloud server, this strategy allows the verifier to begin preparation based on a known prefix even before the edge has finished its entire speculative task. Even in complex scenarios with network jitter or fluctuating acceptance rates, the Fast Verification mechanism ensures a smooth inference pipeline through tighter time-slot scheduling. This computational head-start logic provides PicoSpec with strong "latency immunity" against long-distance  delays.

\subsection{Efficient Separate Rejection Sampling Algorithm}

\begin{figure}[htbp] 
    \centering
    \includegraphics[width=0.9\linewidth]{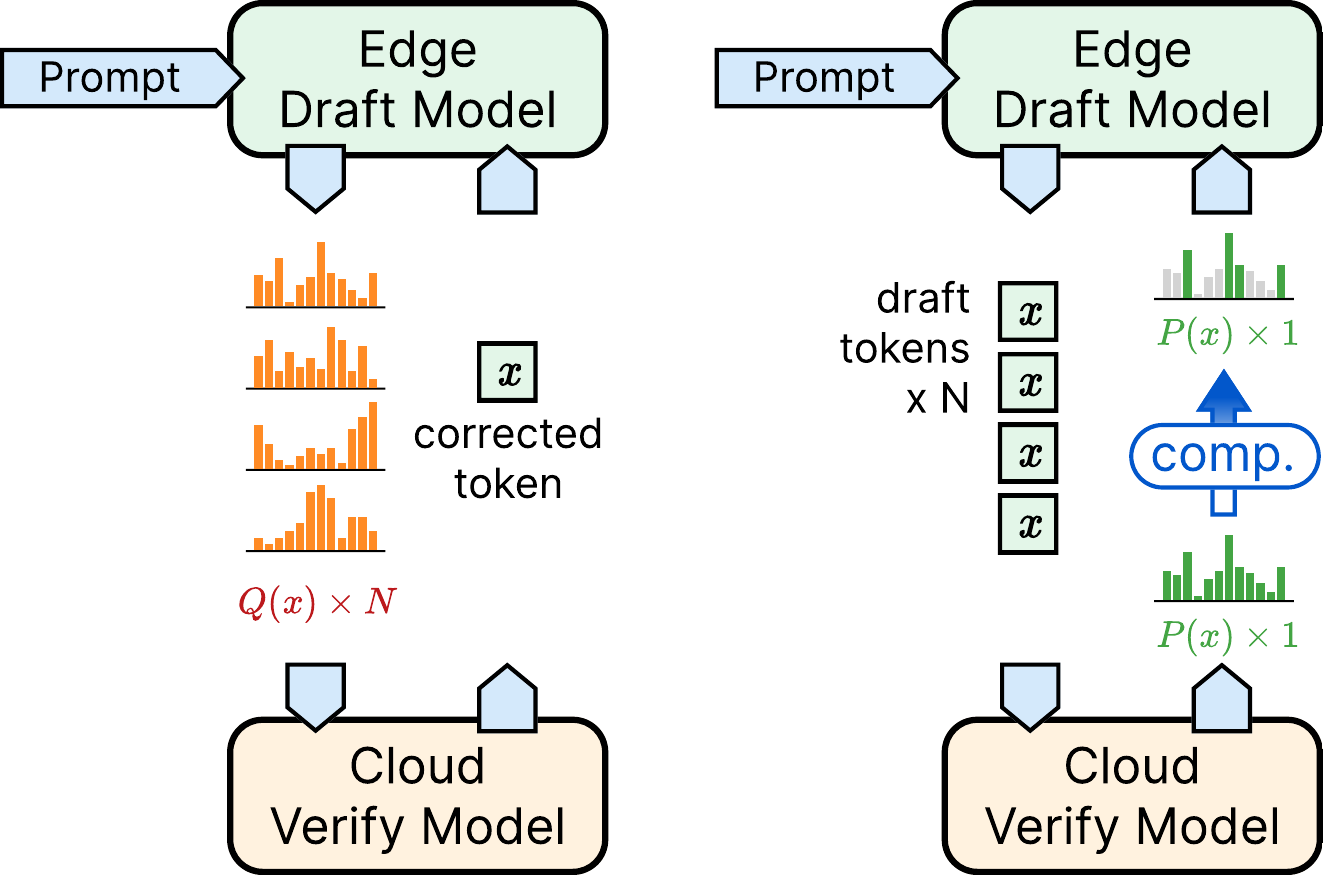}
    \caption{(Left) Vanilla rejection sampling \& (Right) Ours}
    \label{fig:comm} 
\end{figure}

To avoid the communication-overhead trap, PicoSpec uses a specialized algorithm to minimize the amount and the frequency of data exchanged between the edge and the cloud.As shown in Figure \ref{fig:comm}.

In standard distributed setups, transmitting full vocabulary distributions is often too expensive for WAN bandwidth. To solve this by introducing a decoupled resampling method. Unlike traditional distributed speculative decoding that the cloud handles both verification and correction, during each step, instead of full vocabulary distributions, the edge only records and transmits the specific probabilities ($q_1 \dots q_\gamma$) corresponding to these candidates. This reduces the uplink payload to less than 50 bytes per round, significantly lower than the tens of kilobytes required in conventional designs. The cloud LLM performs parallel verification to determine the acceptance of each candidate. If a token is rejected at position $j$, the cloud identifies the divergence and prepares the target distribution $P_j(x)$. Downlink transmission occurs only upon rejection. The edge then autonomously performs local resampling using the logic $norm(\max(0, P_j(x) - Q_j(x)))$ to produce the corrected token.

After verification, return the edge model's vocabulary size $V$, where typically $128,256$ for Llama-3.1 70B and up to $151,936$ for Qwen3. Transmitting such huge data creates significant bandwidth pressure. To further compress the downlink payload, PicoSpec uses a sparse compression strategy on probability data. The cloud extracts only the $K$ components with the highest probabilities and their corresponding indices:

\begin{align}   
    P_{sent} = \{ (id_j, prob_j) \mid j \in \{1, \dots, K\}, \notag \\ 
    prob_j \in \text{TopK}(P_{auth}) \}
\end{align}

Since LLM probability distributions are typically very long-tailed, Top-$K$ pruning can reduce the transmission load by two orders of magnitude while keeping most of the "probability mass". By extracting only the Top-K probability components and their corresponding indices, PicoSpec reduces the downlink payload from $O(V)$ to $O(K)$. For instance, with $V = 128,256$ and $K = 10$, the transmission load per rejected token drops from approximately $500$ KB to less than $100$ bytes—a reduction of over three orders of magnitude. Combined with our zero-copy serialization protocol, the edge can reconstruct tensor views in microseconds. 

Finally, to handle the network jitter and changing edge loads, we include a latency-aware truncation mechanism. In the real world, if generating a draft takes too long, the cloud server will sit idle, wasting its massive computing power. PicoSpec monitors the drafting time in real-time. If the edge is struggling to keep up, the system automatically stops the current speculation, cuts the segment short, and sends it immediately. This dynamic adjustment keeps the pipeline flowing smoothly even when the environment is unstable, masking the latency.

\subsection{Framework Performance Analysis}

To explore the performance boundaries of PicoSpec, we conduct a formal analysis of end-to-end throughput using a probabilistic model.

Define the following parameters: $\gamma$ as the speculative step size; $\alpha$ as the token acceptance rate (the independent probability that a draft token is accepted by the target model); $T_{draft}$ as the total time for the edge to generate $\gamma$ draft tokens; $T_{verify}$ as the time for the cloud to verify $\gamma$ tokens; and $T_{RTT}$ as the network round-trip time, including serialization and deserialization overhead.

In a single speculative cycle, the number of successfully generated tokens, $L$, follows a truncated geometric distribution. The expected generation length $EL$ can be expressed as:

\begin{equation}
   EL = \sum_{k=1}^{\gamma} k \cdot P(L=k) = \frac{1 - \alpha^{\gamma}}{1 - \alpha} 
\end{equation}

As $\alpha \to 1$, $EL \to \gamma$; as $\alpha \to 0$, $EL \to 1$.

In a traditional synchronous stop-and-wait mode, the edge must pause after generating drafts to wait for cloud verification. The end-to-end latency for a single cycle, $L_{sync}$, is the linear sum of the time required for each phase:

\begin{equation}
    L_{sync} = T_{draft} + T_{RTT} + T_{verify}
\end{equation}

In this mode, $T_{RTT}$ and $T_{verify}$ are unavoidable costs on the critical path, making the overall speed significantly lower than purely local inference. The theoretical throughput $R_{sync}$ is defined as:

\begin{equation}
    R_{sync} = \frac{EL}{L_{sync}} = \frac{1 - \alpha^{\gamma}}{(1 - \alpha)(T_{draft} + T_{RTT} + T_{verify})}
\end{equation}

This formula highlights that network latency $T_{RTT}$ severely limits the throughput ceiling in synchronous systems.

PicoSpec breaks these serial dependencies through its asynchronous scheduling. By Parallel Drafting, the verification of step $i$ and the generation of step $i+1$ occur in parallel. However, the actual efficiency of the pipeline depends on the accuracy of the speculation:

\begin{itemize}
\item Case 1: Full Hit ($L = \gamma$). The pre-computation for step $i+1$ is valid, keeping the pipeline full and effectively hiding the network latency. This occurs with probability $P_{hit} = \alpha^{\gamma}$. The amortized time is $T_{hit} = \max(T_{draft}, T_{RTT} + T_{verify})$.

\item Case 2: Miss ($L < \gamma$). Step $i+1$ is generated based on an incorrect context and must be discarded (pipeline flush), resulting in a bubble. The execution effectively reverts to a serial pattern with probability $P_{miss} = 1 - \alpha^{\gamma}$. The time cost is $T_{miss} = T_{sync} = T_{draft} + T_{RTT} + T_{verify}$.
\end{itemize}

Accordingly, the expected amortized time per step for PicoSpec, $ET_{async}$, is calculated as:
\begin{equation}
    E[T_{async}] = P_{hit} \cdot T_{hit} + (1 - P_{hit}) \cdot T_{miss}
\end{equation}
\begin{align}
E[T_{\text{async}}] &= \alpha^{\gamma} \cdot \max(T_{\text{draft}}, T_{\text{RTT}} + T_{\text{verify}}) \notag \\
                    &+ (1 - \alpha^{\gamma})(T_{\text{draft}} + T_{\text{RTT}} + T_{\text{verify}})
\end{align}

The theoretical throughput of PicoSpec is then $R_{async} = {EL} / {ET_{async}}$.

Define the theoretical speedup as $S = R_{async} / R_{sync}$13. In a compute-bound scenario ($T_{draft} \ge T_{RTT} + T_{verify}$) with high model alignment ($\alpha \approx 1$), the performance gain reaches its upper bound:

\begin{equation}
    \lim_{\alpha \to 1} S = \frac{T_{draft} + T_{RTT} + T_{verify}}{T_{draft}} = 1 + \frac{T_{RTT} + T_{verify}}{T_{draft}}
\end{equation}

This theoretical framework leads to two important conclusions:

\begin{itemize}
\item Latency Immunity: Provided $\alpha$ is sufficiently high, PicoSpec can effectively remove $T_{RTT}$ from the bottleneck, allowing collaborative inference speeds to approach the physical limit of local-only draft inference.
\item Robustness Lower Bound: Even in the worst-case scenario ($\alpha \to 0$), PicoSpec adaptively degrades to the performance of the synchronous baseline ($S \to 1$) rather than collapsing due to communication overhead.
\end{itemize}

\section{Evaluation}

\subsection{Setup}

In this paper, using an NVIDIA Jetson AGX as the edge device to represent resource-constrained hardware. The cloud side is supported by a high-performance server equipped with NVIDIA A100 (40GB) GPUs, which handle the heavy computational load of the verifier models. Communication between the edge and the cloud occurs over a Wide Area Network (WAN) characterized by high latency and restricted bandwidth. Our evaluation focuses on two representative model pairs: Qwen 3 0.6B and Qwen 3 32B(using 2 A100 GPUs), Llama 3.2 1B and Llama 3.1 70B(using 4 A100 GPUs). In each pair, the smaller model serves as the edge-side draft model while the larger counterpart acts as the cloud-side verifier. The draft length in speculative decoding is set to 4. These models are tested  across two widely recognized benchmarks: GSM8K for mathematical reasoning and HumanEval for code generation. These datasets allow us to evaluate the framework's performance across different types of logical and structural inference tasks. 

\subsection{Performance Evaluation}

Through compare PicoSpec throughput against two baselines: standard Cloud-Autoregressive (AR), vanilla Speculative decoding and Split Inference. As shown in Table \ref{tab:performance_evaluation}, PicoSpec perform well across all tested model pairs and datasets, achieving up to a 2.9x speedup in high-latency environments. 

Interestingly, vanilla speculative decoding often performs even worse than the AR baseline. For instance, on the QWEN-humaneval task, vanilla speculation only reaches 0.44x (6.24 tokens/s), falling far behind AR’s 14.18 tokens/s. This happens because the stop-and-wait cost of traditional speculation is so high that the network delay completely outweighs any gains from the draft model.For the Split Inference baseline, we deployed the first 3 layers of the model on the NVIDIA Jetson AGX. Chosing this specific configuration because some research\cite{11105796,sung2025memory} indicates that resource-constrained edge devices like the AGX reach an optimal state of computational efficiency and I/O balance when hosting between 2 to 4 layers. However, even with this optimized split, the inference speed remains strictly bound by the network physical limits. However, under the qwen-gsm8k setting, its performance exceeds to vanilla speculative decoding because the lower acceptance rate causes vanilla speculative decoding to block frequently.

\begin{table}[htbp]
\centering
\small 
\setlength{\tabcolsep}{4pt} 
\caption{Transposed Throughput Comparison: Speedup relative to the Autoregressive baseline (tokens/s).}
\label{tab:performance_evaluation}
\resizebox{\columnwidth}{!}{
\begin{tabular}{lcccc} 
\toprule
& \multicolumn{2}{c}{\textbf{Qwen 0.6B \& 32B}} & \multicolumn{2}{c}{\textbf{Llama 1B \& 70B}} \\ 
\cmidrule(lr){2-3} \cmidrule(lr){4-5}
\textbf{Method} & \textbf{GSM8K} & \textbf{HumanEval} & \textbf{GSM8K} & \textbf{HumanEval} \\ 
\midrule
Autoregressive  & 1.00x (13.89) & 1.00x (14.18) & 1.00x (6.87) & 1.00x (6.86) \\
Vanilla Spec.  & 0.58x & 0.44x & 1.35x & 1.45x \\
Split Inf. & 0.54x & 0.53x & 0.63x & 0.66x \\
\textbf{PicoSpec (Ours)}  & \textbf{1.45x} & \textbf{1.13x} & \textbf{2.51x} & \textbf{2.90x} \\
\midrule
Avg. Acpt. L.(Ours) & 2.50 & 1.88 & 4.29 & 4.90 \\
\bottomrule
\end{tabular}
}
\end{table}

For the Qwen (0.6B \& 32B) pair, PicoSpec improves throughput from 13.89 to 20.19 tokens/s (+45.3\%) on the GSM8K dataset, and from 14.18 to 16.04 tokens/s (+13.1\%) on HumanEval. The performance leap is even more pronounced for the Llama (1B \& 70B) pair, where throughput surges from approximately 6.8 tokens/s to 17.22 tokens/s (+150.6\%) on GSM8K and 19.88 tokens/s (+189.8\%) on HumanEval.

The speedup is notably higher for the Llama pair than for the Qwen pair. This is because the larger Llama-70B model incurs a significantly higher $T_{verify}$ compared to the 32B Qwen model. Both $T_{draft}$ and $T_{verify}$ are delays that cannot be ignored in the edge-cloud environment. In the AR and vanilla speculation baseline, this larger cloud latency directly increases the cost of every token. Conversely, PicoSpec provides more room to hide the larger $T_{verify}$ and RTT within the continuous drafting cycle of the edge device. This proves PicoSpec's "latency immunity", where collaborative inference speeds approach the physical drafting limit of local execution regardless of remote delays.

In addition, it can be observed that gsm8k performs better on the Qwen model pair, while humaneval performs better on the llama model pair. This is due to different models are good at the different tasks. It can be seen from the table of accept lengths for different experimental settings, PicoSpec's performance is positively correlated with the token acceptance length. Scenarios with higher acceptance rates—such as the Llama pair on HumanEval reaching a peak of 4.90 tokens—allow the asynchronous pipeline to maintain a full hit state more consistently. In such cases, the system can more effectively mask the network RTT, minimizing pipeline rollbacks and maximizing the throughput of our parallel drafting mechanism.

\subsection{Ablation Experiment}

To investigate the contribution of each core component in PicoSpec, we conduct an extensive ablation study by isolating three key mechanisms: asynchronous pipelining (Para-draft),  Fast Verification(Fast-verify), and separate rejection sampling (Split-rej). The results, detailed in Table\ref{tab:ablation_qwen} \& \ref{tab:ablation_llama}, demonstrate that the synergy of these components is essential for maintaining high performance in high-latency environments.

The removal of the asynchronous pipeline (w/o Para-draft) results in the most significant performance degradation. Without this mechanism, the system reverts to a traditional serial stop-and-wait paradigm where the edge-side NVIDIA Jetson AGX must remain idle during the entire $T_{RTT} + T_{verify}$ interval.  For the Llama-gsm8k setting, the throughput drops from 17.22 tokens/s to 12.51 tokens/s. This confirms that decoupling computation from communication is the primary driver for masking network latency.

The Separate Rejection Sampling algorithm is important for minimizing communication overhead. When this component is disabled (w/o Split-rej), the system is forced to transmit huge probability distributions instead of specific probabilities and compressed results. This leads to a massive surge in $T_{verify}$ due to increased bandwidth pressure and CPU deserialization costs. Specifically, in the Llama-gsm8k scenario, $T_{verify}$ jumps from 166.46ms to 283.79ms, highlighting the necessity of communication-efficient sampling for distributed speculative decoding. 

The  Fast Verification strategy aims to further eliminate residual bubble in the pipeline by initiating cloud preparation before the entire draft batch is finalized.  Removing this feature (w/o Fast-verify) consistently increases the Time Per Output Token (TPOT). For instance, in the QWEN-gsm8k case, the throughput decreases from 20.19 to 17.87 tokens/s. This illustrates that  Fast Verification provides an essential computational head-start, ensuring a smooth inference flow even when network jitter occur or low acceptance rate.

\begin{table}[h]
\centering
\caption{Ablation results for the Qwen model pair (0.6B and 32B).}
\label{tab:ablation_qwen}
\small
\resizebox{\columnwidth}{!}{
\begin{tabular}{lccccc}
\toprule
\textbf{Configuration} & \textbf{Thr. $\uparrow$} & \textbf{TTFT $\downarrow$} & \textbf{TPOT $\downarrow$} & \textbf{$T_{verify} \downarrow$} & \textbf{$T_{draft} \downarrow$} \\ 
& (tokens/s) & (ms) & (ms) & (ms) & (ms) \\
\midrule
\rowcolor[gray]{.95} \multicolumn{6}{l}{\textit{Dataset: gsm8k}} \\
w/o Para-draft & 11.92 & 148.96 & 251.76 & 119.81 & 123.91 \\
w/o Fast-verify & 17.87 & 150.32 & 166.56 & 106.56 & 97.30 \\
w/o Split-rej & 12.62 & 154.68 & 233.30 & 166.98 & 100.96 \\
\textbf{Ours (Full)} & \textbf{20.19} & \textbf{143.99} & \textbf{148.84} & \textbf{87.35} & \textbf{97.46} \\
\midrule
\rowcolor[gray]{.95} \multicolumn{6}{l}{\textit{Dataset: humaneval}} \\
w/o Parallel & 10.25 & 177.90 & 252.92 & 119.03 & 124.08 \\
w/o Fast-verify & 13.86 & 164.84 & 187.07 & 109.14 & 97.16 \\
w/o Split-rej & 10.23 & 164.82 & 251.73 & 168.27 & 99.98 \\
\textbf{Ours (Full)} & \textbf{16.04} & \textbf{155.11} & \textbf{160.88} & \textbf{78.04} & \textbf{97.17} \\
\bottomrule
\end{tabular}
}
\end{table}

\begin{table}[h]
\centering
\caption{Ablation results for the Llama model pair (1B and 70B).}
\label{tab:ablation_llama}
\small
\resizebox{\columnwidth}{!}{
\begin{tabular}{lccccc}
\toprule
\textbf{Configuration} & \textbf{Thr. $\uparrow$} & \textbf{TTFT $\downarrow$} & \textbf{TPOT $\downarrow$} & \textbf{$T_{verify} \downarrow$} & \textbf{$T_{draft} \downarrow$} \\ 
& (tokens/s) & (ms) & (ms) & (ms) & (ms) \\
\midrule
\rowcolor[gray]{.95} \multicolumn{6}{l}{\textit{Dataset: gsm8k}} \\
w/o Parallel & 12.51 & 305.46 & 390.35 & 194.68 & 190.56 \\
w/o Fast-verify & 16.64 & 297.40 & 287.74 & 189.32 & 170.93 \\
w/o Split-rej & 12.12 & 303.17 & 389.76 & 283.79 & 173.69 \\
\textbf{Ours (Full)} & \textbf{17.22} & \textbf{295.58} & \textbf{272.89} & \textbf{166.46} & \textbf{171.05} \\
\midrule
\rowcolor[gray]{.95} \multicolumn{6}{l}{\textit{Dataset: humaneval}} \\
w/o Parallel & 13.60 & 342.68 & 380.89 & 186.60 & 189.68 \\
w/o Fast-verify & 19.67 & 346.16 & 263.94 & 182.32 & 171.64 \\
w/o Split-rej & 14.29 & 344.35 & 366.93 & 281.84 & 173.60 \\
\textbf{Ours (Full)} & \textbf{19.88} & \textbf{369.02} & \textbf{272.93} & \textbf{182.96} & \textbf{172.41} \\
\bottomrule
\end{tabular}
}
\end{table}

\subsection{Parameter Sensitivity}

This section investigates how the draft generation length ($n$) affects PicoSpec's end-to-end performance, using the QWEN-gsm8k setting as a representative case study. As shown in Table \ref{tab:draft_len_selection}, the choice of $n$ involves a critical trade-off between speculative gains and local computational overhead. We observe that while the average acceptance length grows steadily from 2.03 to 3.10 as $n$ increases, the system throughput remains remarkably stable within the $n=3$ to $n=5$ range, staying above 18 tokens/s. This show that PicoSpec is robust and does not require extremely precise parameter tuning to achieve significant speedups.

The throughput reaches its peak of 20.19 tokens/s at $n=4$. This represents the draft generation time ($T_{draft} \approx 97ms$) effectively overlaps with the sum of network RTT and cloud verification time, successfully masking the communication latency. However, as $n$ grows beyond this point, the performance begins to decline. This is because the local drafting cost starts to escalate—rising from 82.11ms at $n=3$ to 145.40ms at $n=6$—which eventually outweighs the benefits of a higher acceptance rate. Consequently, the Time Per Output Token (TPOT) increases significantly, and the system moves away from its ideal non-blocking state. Based on these observations, we select $n=4$ as the default configuration for our framework.

\begin{table}[h]
\centering
\caption{Performance comparison under different draft generation lengths.}
\label{tab:draft_len_selection}
\small
\resizebox{\columnwidth}{!}{
\begin{tabular}{ccccccc}
\toprule
\textbf{Draft Len} & \textbf{Thr. $\uparrow$} & \textbf{Acc. Len $\uparrow$} & \textbf{TTFT $\downarrow$} & \textbf{TPOT $\downarrow$} & \textbf{$T_{verify} \downarrow$} & \textbf{$T_{draft} \downarrow$} \\ 
($n$) & (tokens/s) & & (ms) & (ms) & (ms) & (ms) \\
\midrule
3 & 19.06 & 2.03 & 146.55 & 129.15 & 82.11 & 73.41 \\
\textbf{4} & \textbf{20.19} & 2.50 & 143.99 & 148.84 & 87.35 & 97.46 \\
5 & 18.12 & 2.84 & 143.39 & 189.17 & 88.76 & 121.40 \\
6 & 17.12 & 3.10 & 143.26 & 219.28 & 88.07 & 145.40 \\
\bottomrule
\end{tabular}
}
\end{table}

\section{Conclusion}
This paper present PicoSpec, a plug-and-play framework designed to overcome the communication-computation mismatch in Edge-Cloud speculative decoding. By decoupling edge-side drafting from cloud-side verification through an asynchronous pipeline, PicoSpec effectively masks round-trip times without requiring model modifications or retraining. Experiments conducted using an NVIDIA Jetson AGX and an A100 cluster demonstrate that our framework achieves significant "latency immunity". For large models like Llama-70B, the method can run up to 2.9 times faster than cloud-only autoregressive models. While PicoSpec have strong performance, its speedup remains naturally bounded by the alignment between the draft models and target models. But, the system can ensures robustness by degrading to a synchronous baseline in the worst-case.

\bibliographystyle{named}
\bibliography{ijcai26}

@article{chen2023accelerating,
  title={Accelerating large language model decoding with speculative sampling},
  author={Chen, Charlie and Borgeaud, Sebastian and Irving, Geoffrey and Lespiau, Jean-Baptiste and Sifre, Laurent and Jumper, John},
  journal={arXiv preprint arXiv:2302.01318},
  year={2023}
}

@InProceedings{pmlr-v235-cai24b,
  title = 	 {Medusa: Simple {LLM} Inference Acceleration Framework with Multiple Decoding Heads},
  author =       {Cai, Tianle and Li, Yuhong and Geng, Zhengyang and Peng, Hongwu and Lee, Jason D. and Chen, Deming and Dao, Tri},
  booktitle = 	 {Proceedings of the 41st International Conference on Machine Learning},
  pages = 	 {5209--5235},
  year = 	 {2024},
  volume = 	 {235},
  series = 	 {Proceedings of Machine Learning Research},
  month = 	 {21--27 Jul},
}

@inproceedings{10.5555/3692070.3693232,
author = {Li, Yuhui and Wei, Fangyun and Zhang, Chao and Zhang, Hongyang},
title = {EAGLE: speculative sampling requires rethinking feature uncertainty},
year = {2024},
publisher = {JMLR.org},
booktitle = {Proceedings of the 41st International Conference on Machine Learning},
articleno = {1162},
numpages = {14},
location = {Vienna, Austria},
series = {ICML'24}
}

@article{zhang2025swiftspec,
  title={SwiftSpec: Ultra-Low Latency LLM Decoding by Scaling Asynchronous Speculative Decoding},
  author={Zhang, Ziyi and Jiang, Ziheng and Jiang, Chengquan and Yu, Menghan and Zheng, Size and Lin, Haibin and Hoffmann, Henry and Liu, Xin},
  journal={arXiv preprint arXiv:2506.11309},
  year={2025}
}

@article{gao2025collaborative,
  title={Collaborative speculative inference for efficient llm inference serving},
  author={Gao, Luyao and Liu, Jianchun and Xu, Hongli and Zhang, Xichong and Liao, Yunming and Huang, Liusheng},
  journal={arXiv preprint arXiv:2503.10325},
  year={2025}
}

@article{zhang2024edgeshard,
  title={Edgeshard: Efficient llm inference via collaborative edge computing},
  author={Zhang, Mingjin and Shen, Xiaoming and Cao, Jiannong and Cui, Zeyang and Jiang, Shan},
  journal={IEEE Internet of Things Journal},
  year={2024},
  publisher={IEEE}
}

@inproceedings{ye2025jupiter,
  title={Jupiter: Fast and resource-efficient collaborative inference of generative llms on edge devices},
  author={Ye, Shengyuan and Ouyang, Bei and Zeng, Liekang and Qian, Tianyi and Chu, Xiaowen and Tang, Jian and Chen, Xu},
  booktitle={IEEE INFOCOM 2025-IEEE Conference on Computer Communications},
  pages={1--10},
  year={2025},
  organization={IEEE}
}

@inproceedings{ma2025multi,
  title={Multi-tier multi-node scheduling of llm for collaborative ai computing},
  author={Ma, Mulei and Gong, Chenyu and Zeng, Liekang and Yang, Yang},
  booktitle={IEEE INFOCOM 2025-IEEE Conference on Computer Communications},
  pages={1--10},
  year={2025},
  organization={IEEE}
}

@inproceedings{ding2024hybrid,
title={Hybrid {LLM}: Cost-Efficient and Quality-Aware Query Routing},
author={Dujian Ding and Ankur Mallick and Chi Wang and Robert Sim and Subhabrata Mukherjee and Victor R{\"u}hle and Laks V. S. Lakshmanan and Ahmed Hassan Awadallah},
booktitle={The Twelfth International Conference on Learning Representations},
year={2024},
url={https://openreview.net/forum?id=02f3mUtqnM}
}

@inproceedings{hao2024hybrid,
  title={Hybrid slm and llm for edge-cloud collaborative inference},
  author={Hao, Zixu and Jiang, Huiqiang and Jiang, Shiqi and Ren, Ju and Cao, Ting},
  booktitle={Proceedings of the Workshop on Edge and Mobile Foundation Models},
  pages={36--41},
  year={2024}
}

@article{xu2024edgellm,
  title={Edgellm: Fast on-device llm inference with speculative decoding},
  author={Xu, Daliang and Yin, Wangsong and Zhang, Hao and Jin, Xin and Zhang, Ying and Wei, Shiyun and Xu, Mengwei and Liu, Xuanzhe},
  journal={IEEE Transactions on Mobile Computing},
  year={2024},
  publisher={IEEE}
}

@article{venkatesha2025fast,
  title={Fast and Cost-effective Speculative Edge-Cloud Decoding with Early Exits},
  author={Venkatesha, Yeshwanth and Kundu, Souvik and Panda, Priyadarshini},
  journal={arXiv preprint arXiv:2505.21594},
  year={2025}
}

@article{xie2025novel,
  title={A Novel Hat-Shaped Device-Cloud Collaborative Inference Framework for Large Language Models},
  author={Xie, Zuan and Xu, Yang and Xu, Hongli and Liao, Yunming and Yao, Zhiwei},
  journal={arXiv preprint arXiv:2503.18989},
  year={2025}
}

@inproceedings{ning2025dssd,
  title={DSSD: Efficient Edge-Device Deployment and Collaborative Inference via Distributed Split Speculative Decoding},
  author={Ning, Jiahong and Zheng, Ce and Yang, Tingting},
  booktitle={ICML 2025 Workshop on Machine Learning for Wireless Communication and Networks (ML4Wireless)},
  year={2025}
}

@article{yu2025dsd,
  title={DSD: A Distributed Speculative Decoding Solution for Edge-Cloud Agile Large Model Serving},
  author={Yu, Fengze and Li, Leshu and McDanel, Brad and Zhang, Saiqian},
  journal={arXiv preprint arXiv:2511.21669},
  year={2025}
}

@inproceedings{li2025sled,
  title={Sled: A speculative llm decoding framework for efficient edge serving},
  author={Li, Xiangchen and Spatharakis, Dimitrios and Ghafouri, Saeid and Fan, Jiakun and Vandierendonck, Hans and John, Deepu and Ji, Bo and Nikolopoulos, Dimitrios S},
  booktitle={Proceedings of the Tenth ACM/IEEE Symposium on Edge Computing},
  pages={1--8},
  year={2025}
}

@article{10.1145/3719664,
author = {Zheng, Yue and Chen, Yuhao and Qian, Bin and Shi, Xiufang and Shu, Yuanchao and Chen, Jiming},
title = {A Review on Edge Large Language Models: Design, Execution, and Applications},
year = {2025},
issue_date = {August 2025},
publisher = {Association for Computing Machinery},
address = {New York, NY, USA},
volume = {57},
number = {8},
issn = {0360-0300},
url = {https://doi.org/10.1145/3719664},
doi = {10.1145/3719664},
month = mar,
articleno = {209},
numpages = {35},
}

@article{wang2024comprehensive,
author = {Wang, Fali and Zhang, Zhiwei and Zhang, Xianren and Wu, Zongyu and Mo, TzuHao and Lu, Qiuhao and Wang, Wanjing and Li, Rui and Xu, Junjie and Tang, Xianfeng and He, Qi and Ma, Yao and Huang, Ming and Wang, Suhang},
title = {A Comprehensive Survey of Small Language Models in the Era of Large Language Models: Techniques, Enhancements, Applications, Collaboration with LLMs, and Trustworthiness},
year = {2025},
publisher = {Association for Computing Machinery},
address = {New York, NY, USA},
issn = {2157-6904},
url = {https://doi.org/10.1145/3768165},
doi = {10.1145/3768165},
note = {Just Accepted},
journal = {ACM Trans. Intell. Syst. Technol.},
month = sep
}

@inproceedings{liu2025pearl,
title={{PEARL}: Parallel Speculative Decoding with Adaptive Draft Length},
author={Tianyu Liu and Yun Li and Qitan Lv and Kai Liu and Jianchen Zhu and Winston Hu and Xiao Sun},
booktitle={The Thirteenth International Conference on Learning Representations},
year={2025},
url={https://openreview.net/forum?id=QOXrVMiHGK}
}

@INPROCEEDINGS{11105796,
  author={Arya, Mayank and Simmhan, Yogesh},
  booktitle={2025 IEEE International Parallel and Distributed Processing Symposium Workshops (IPDPSW)}, 
  title={Understanding the Performance and Power of LLM Inferencing on Edge Accelerators}, 
  year={2025},
  pages={1108-1111},
  doi={10.1109/IPDPSW66978.2025.00173}}

@article{sung2025memory,
  title={Memory-and Latency-Constrained Inference of Large Language Models via Adaptive Split Computing},
  author={Sung, Mingyu and Palakonda, Vikas and Im, Suhwan and Moon, Sunghwan and Kim, Il-Min and Yun, Sangseok and Kang, Jae-Mo},
  journal={arXiv preprint arXiv:2511.04002},
  year={2025}
}

\end{document}